\documentstyle[11pt,a4,cite]{article}
\def\ga{\alpha}

\def\ge{\epsilon}
\def\gg{\gamma}
\def\gd{\delta}

\def\gm{\mu}
\def\gn{\nu}
\def\gp{\pi}
\def\gP{\Pi}

\def\gs{\sigma}

\def\gL{\Lambda}
\def\gt{\theta}

\def\gz{\zeta}
\def\delp{\partial_+}
\def\delm{\partial_-}

\def\delrlp{\stackrel {\leftrightarrow} {\partial_+}}
\def\delrlm{\stackrel {\leftrightarrow} {\partial_-}} 
\def\part{\partial}
\def\hlf{\frac{1}{2}}
\def\A0{A^{+}_0}
\def\psip{\psi_+}
\def\psin{\psi_-}
\def\psipd{\psi^{\dagger}_+}
\def\psind{\psi^{\dagger}_-}

\def\psid{\psi^{\dag}}

\def\xpl{x^{+}}
\def\xmin{x^{-}}
\def\ymin{y^{-}}
\newcommand{\nc}{\newcommand}
\nc{\intgl}{\int\limits_{-L}^{+L}\!{{dx^-}\over\!2}}
\nc{\intgly}{\int\limits_{-L}^{+L}\!{{dy^-}\over\!2}}
\nc{\zmint}{\int\limits_{-L}^{+L}\!{{dx^-}\over{\!2L}}}
\def\beq{\begin{equation}}
\def\eeq{\end{equation}}
\def\bea{\begin{eqnarray}}
\def\eea{\end{eqnarray}}
\begin{document}
\title{ Large Gauge Transformations and the Light-Front Vacuum Structure}
\medskip
\author{{\sl  L$\!\!$'ubom\'{\i}r Martinovi\v c} \\
Institute of Physics, Slovak Academy of Sciences \\
D\'ubravsk\'a cesta 9, 842 28 Bratislava, Slovakia \thanks{permanent address}\\
and\\
International Institute of Theoretical and Applied Physics\\
Iowa State University, Ames, Iowa 50011, USA}
\date{}
\maketitle
\begin{abstract}

A residual gauge symmetry, exhibited by light-front gauge theories 
quantized in a finite volume, is analyzed at the quantum level. Unitary 
operators, which implement the symmetry, transform the trivial Fock vacuum 
into an infinite set of degenerate coherent-state vacua. A fermionic component 
of the vacuum emerges naturally without the need to introduce a Dirac sea. The 
vacuum degeneracy along with the derivation of the theta-vacuum is discussed 
within the massive Schwinger model. A possible generalization 
of the approach to more realistic gauge field theories is suggested.  

\end{abstract}

\section{Introduction}

Hamiltonian quantum field theory formulated in the light front (space-time and  
field) variables \cite{Dir,Sus,Leutw,LKS,Rohr} has often been considered as a 
conceptually very attractive theoretical scheme. Vacuum aspects of the dynamics 
seem to simplify remarkably (Fock vacuum is to a very good approximation an  
eigenstate of the full Hamiltonian) at the same time causing problems with  
understanding chiral properties, vacuum degeneracy and symmetry breaking  
phenomena. For example, it is not clear how one could reproduce the axial 
anomaly and the chiral condensate \cite{Schw,LSw} in the light-front Schwinger 
model. These and related difficulties \cite{Dave} have been usually explained  
by the ``peculiarities'' of the quantization on the characteristic  
surface $x^+=0$ \cite{McC,Rgf}. 

In the present work, one of the so far missing components of the light-front 
(LF) gauge field theory, namely the non-trivial vacuum structure, is found to 
be directly related to a residual ``large" gauge symmetry present in the 
finite-box formulation \cite{Mant} of the theory. The general idea is of 
course not new. Gauge transformations with non-trivial topological properties 
have been shown to be responsible for the vacuum degeneracy in \cite{tH,CDG,JR,
RotheS,Stroc,IsoM,Adam}, e.g.. Their role has been studied also in the 
light-front literature \cite{Franke,Rgb,Rgf,KPP,Alex}. 

The novel feature in our approach is the quantum-mechanical implementation of 
large gauge transformations by unitary operators in the context of the 
``trivial'' non-perturbative Fock vacuum of the LF field theory. The unitary 
operators act on the fields as well as on states in Hilbert space. As a 
consequence, the ``trivial" LF Fock vacuum transforms into an infinite set of 
non-trivial vacua. They are basically coherent states of both the dynamical 
gauge-field zero mode and an effective boson field bilinear in dynamical fermi 
field operators. The multiple vacua can be superimposed to form a unique gauge 
invariant vacuum. This will be shown with the example of the (massive) 
Schwinger model, which is known to exhibit in a tractable form many of  
non-perturbative features expected in QCD. We will argue however that the 
mechanism could in principle work also for more complicated gauge theories.

\section{LF Quantization of the Massive Schwinger Model}

Due to specific light-front constraints, it is inevitable to adopt the  
Dirac-Bergmann (DB) \cite{DB} or other similar method to properly quantize  
the LF massive Schwinger model \cite{Hara,LMlong}. Here we quote only those 
results of the DB analysis which are relevant for our approach to the vacuum 
problem. 
 
In terms of the LF variables, the Lagrangian density of the two-dimensional 
spinor field $\Psi$ of mass $m$ interacting with the gauge field $A^\gm$ takes 
the form
\bea  
{\cal L}_{LF} = i\psipd\delrlp\psip + i\psind\delrlm\psin +  
\hlf(\delp A^{+} - \delm A^{-})^2  - \nonumber \\- m(\psipd\psin + 
\psind\psip) - {e \over 2}j^{+}A^{-} - {e \over 2}j^{-}A^{+} .
\label{lflagr}
\eea  
We choose $x^+ = x^0 + x^1$ and $\xmin = x^0 - x^1$ as the LF time and space 
variable, correspondingly. The dynamical ($\psip$) and dependent ($\psin$) 
projections of the fermi field are defined as $\psi_{\pm} = \gL_{\pm}\Psi$,   
where $\gL_{\pm}=\hlf\gg^0\gg^{\pm}, \gg^{\pm}=\gg^0 \pm \gg^1, \gg^0 =\gs^1, 
\gg^1 = i\gs^2$ and $\gs^1, \gs^2$ are the Pauli matrices. At the quantum  
level, the vector current will be represented by normal-ordered product of the 
fermi operators, $j^{\pm}=2:\psid_{\pm}\psi_{\pm}:$.

A suitable finite-interval formulation of the model is achieved by imposing
the restriction $-L \le x^- \le L$ and by prescribing antiperiodic boundary 
conditions for the fermion field and periodic ones for the gauge field. The 
latter imply a decomposition of the gauge field into the zero-mode (ZM) part  
$A_0^{\gm}$ and the part $A^{\gm}_{n}$ containing only normal Fourier modes. 
We will work in the usual gauge $A^+_n=0, A^-_0=0$, which completely 
eliminates gauge freedom with respect to {\it small} gauge transformations. 
In a finite volume with periodic gauge field, the ZM $\A0$  becomes a physical 
variable \cite{Franke,Mant,HH,Lenz91,Rgb} since it cannot be gauged away. 
In quantum theory, it satisfies the commutation relation
\beq
\left[\A0(\xpl),\gP_{\A0}(\xpl)\right]  =  {i\over{L}},
\label{zmcr}
\eeq 
where $\gP_{\A0}=\delp\A0$ is the momentum conjugate to $\A0$. The DB  
procedure yields the anticommutator for the independent fermi field component  
\beq
\{\psip(\xmin,\xpl),\psipd(\ymin,\xpl)\}  =  \hlf \gL^+ \gd_{a}(\xmin-\ymin)
\label{acr}
\eeq 
with the antiperiodic delta function $\gd_a(\xmin-\ymin)$ \cite{AC} being   
regularized by a LF momentum cutoff $N$. The fermi-field Fock operators are   
defined by   
\beq
\psip(\xmin) = {1 \over{\sqrt{2L}}} \left(\matrix{0 \cr 1}\right)
\!\sum_{n=\hlf}^{N} \! \left(
b_ne^{-{i \over 2}k^+_n\xmin} + d_n^{\dagger}e^{{i \over 2}k^+_n\xmin}\right),
\label{fermexp}
\eeq
\beq
\{b_n,b^{\dagger}_{n^{\prime}}\} = \{d_n,d^{\dagger}_{n^{\prime}}\} = 
\gd_{n,n^{\prime}},\;\;n=\hlf,{3 \over 2},\dots,\;k_n^+ = {2\gp \over L}n.
\eeq
While the LF momentum operator $P^+$ depends only on $\psip$, the gauge 
invariant (see below) LF Hamiltonian of the model is expressed 
in terms of the both unconstrained variables $\psip$ and $\A0$ as 
\bea 
P^- = L\gP^2_{\A0} - {e^2 \over 4}\intgl\intgly j^+(\xmin)
{\cal G}_2
(\xmin - \ymin) j^+(\ymin) + \nonumber \\ 
+ m^2\intgl\intgly\left[\psid(\xmin){\cal G}_a(\xmin - \ymin;\A0)\psi(\ymin) 
+ h.c. \right] . 
\label{lfham}
\eea   
The Green's functions  
\beq
{\cal G}_2(\xmin - \ymin) = \!-{4\over{L}}\sum_{m=1}^{M}
{1 \over{{p^+_m}^2}}\left(e^{-{i\over {2}}p^+_m(\xmin - \ymin)} + 
e^{{i\over{2}}p^+_m(\xmin - \ymin)} 
\right),\;p_m^+ = {2\gp \over L}m,
\eeq
\beq
{\cal G}_a(\xmin\!-\ymin;\A0) = {1 \over {4i}}\left[\ge(\xmin \!- \ymin) + 
i \tan({{eL}\over{2}}\A0)\right]\!\exp{\left(-{ie\over 2}(\xmin \!- \ymin)\A0
\right)} 
\eeq
have been used to eliminate the constrained variables $A^{-}_n$ and $\psin$, 
respectively, with $\ge(\xmin)$ being twice the sign function, 
$\delm \ge(\xmin) = 2\gd_a(\xmin)$. 

The final consequence of the DB analysis is the condition (a first-class 
constraint) of electric neutrality of the physical states,   
$Q\vert phys \rangle = 0.$ 

\section{Large Gauge Transformations and Theta-Vacuum}

It is well known that gauge theories quantized in a finite volume exhibit  
an extra symmetry not explicitly present in the continuum approach \cite{Mant,
Lenz91,Rgb,IsoM,Lenzqm,Lenzax}. In the LF formulation, the corresponding gauge 
function is linear in $\xmin$ with a coefficient, given by a specific  
combination of constants. These simple properties follow from the requirement 
to maintain boundary conditions for the gauge and matter field, respectively. 
The above symmetry is the finite-box analogue \cite{Franke,Rgb} of topological 
transformations familiar from the continuum formulation. Note that in the LF 
theory they are restricted to the $+$ gauge field component even in $3+1$ 
dimensions. This simplifies their implementation at the quantum level. 

For the considered $U(1)$ theory, the corresponding gauge function has the form 
$\gL_\gn = {\gp\over L}\gn \xmin$, is non-vanishing at $\pm L$ and defines a 
winding number $\gn$: 
\beq
\gL_\gn(L) - \gL_\gn(-L) = 2\gp\gn,\;\;\gn \in Z. 
\eeq 
Thus, the residual gauge symmetry of the Hamiltonian (\ref{lfham}) is 
\cite{Hara}  
\beq
\A0 \rightarrow \A0 - {2\gp\over{eL}}\gn,\;\; 
\psip(\xmin) \rightarrow e^{i{\gp \over L}\gn\xmin}\psip
(\xmin). 
\label{zmshift}
\eeq

Let us discuss the ZM part of the symmetry first. At the quantum 
level, it is convenient to work with the rescaled ZM operators \cite{AD} 
$\hat{\gz}$ and $\hat{\gp}_0$ : 
\beq
\A0 = {2\gp \over{eL}}\hat{\gz},\;\;\gP_{\A0} = {e \over{2\gp}}\hat{\gp}_0,\;\; 
\left[\hat{\gz},\hat{\gp}_0\right] = i .
\label{qmcr}
\eeq   
Note that the box length dropped out from the ZM commutator. The shift    
transformation of $\A0$ is for $\gn=1$ implemented by the  
unitary operator $\hat{Z}_1$:
\beq
\hat{\gz} \rightarrow \hat{Z}_1 \hat{\gz} \hat{Z}^{\dagger}_1 = \hat{\gz} - 1, 
\;\; \hat{Z}_1 = \exp(-i\hat{\gp}_0) .
\eeq
The transformation of the ZM operator $\hat{\gz}$ is accompanied by 
the corresponding transformation of the vacuum state (see e.g. \cite{IZ} for   
a related example), which we define by $a_0\vert 0 \rangle = 0$. 
$a_0(a_0^\dagger)$ is the usual annihilation (creation) operator of a boson 
quantum:
\beq
a_0 = {1 \over{\sqrt{2}}}\left(\hat{\gz} + i\hat{\gp}_0\right),\;\;
a_0^{\dagger} = {1 \over{\sqrt{2}}}\left(\hat{\gz} - i\hat{\gp}_0\right),\;\;
\left[a_0,a^{\dagger}_0\right] = 1. 
\eeq
$\hat{Z}_1$ is essentially the displacement operator $\hat{D}(\ga)$  
of the theory of coherent states \cite{Glau,Zhang}. In our case, the 
complex parameter $\ga$ is replaced by the integer $\gn$ and the operator  
$\hat{Z}(\gn) = (\hat{Z}_1)^\gn$ takes the form 
\beq
\hat{Z}(\gn) = \exp\left[{\gn\over{\sqrt{2}}}\left(a^{\dagger}_0 - a_0\right)
\right]
\eeq
with the properties
\beq
\hat{Z}(\gn)a_0\hat{Z}^{\dagger}(\gn) = a_0 - {\gn\over{\sqrt{2}}},\;\;\; 
a_0\vert \gn;z \rangle = {\gn\over{\sqrt{2}}}\vert \gn;z \rangle,\;\;\;\vert 
\gn;z \rangle \equiv \hat{Z}(\gn)\vert 0 \rangle .
\label{FCS}
\eeq   
The transformed (displaced) vacuum expressed in terms of the harmonic 
oscillator Fock states $\vert n \rangle$ and the corresponding amplitudes 
$C_n$  \cite{Glau,Zhang}
\beq
\vert \gn;z \rangle =  \sum_{n=0}^{\infty}C_n(\gn)\vert n \rangle
\eeq
can be understood as describing the condensate of zero-mode gauge bosons. 

Alternatively, one may consider the problem in quantum mechanical coordinate  
representation, where $\hat{\gp}_0 = -i{d\over{d\gz}}$ and the vacuum 
wavefunction $\psi_0(\gz)$ transforms as  
\beq 
\psi_0(\gz) = \gp^{-{1\over{4}}}\exp(-{\hlf}\gz^2)  
\rightarrow \psi_{\gn}(\gz) = \exp(-\gn {d\over{d\gz}})
\psi_0(\gz)\ = \gp^{-{1\over{4}}}\exp(-{\hlf}(\gz-\gn)^2).
\eeq   
The ZM kinetic energy term of the LF Hamiltonian (\ref{lfham}) is given by   
\beq
P^-_{0} = -\hlf{{e^2L}\over{2\gp^2}} {d^2\over{d\gz^2}}. 
\eeq
Usually, a Schr\"odinger equation with the above $P_0^-$ (or its equal-time 
counterpart) is invoked to find 
the vacuum energy and the corresponding wave functions subject to a  
periodicity condition at the boundaries of the fundamental domain $0 \le \gz 
\le 1$ \cite{Mant,KPP,ADQED}. Here we are led by simple symmetry 
arguments to consider instead of the lowest-energy eigenfunction of 
$P^-_{0}\sim \hat{\gp}^2_0$ the eigenstates of $a_0$ with a non-vanishing 
eigenvalue $\gn$ -- the ZM coherent states. The corresponding LF energy    
\beq
E_0 = \int\limits_{-\infty}^{+\infty}d\gz\psi_{\gn}(\gz)P^-_{0}\psi_{\gn}
(\gz) = {e^2L\over{8\gp^2}} \;,
\label{zmvacen}
\eeq 
is independent of $\gn$, thus the infinite set of vacuum states  
$\psi_\gn(\gz),\;\gn \in Z$, is degenerate in the LF energy. In addition, they  
are not invariant under $\hat{Z}_1$, 
\beq
\hat{Z}_1\psi_{\gn}(\gz) = \psi_{\gn + 1}(\gz)   
\label{psishift}
\eeq
and those $\psi_\gn(\gz)$ which differ by unity in the value of $\gn$ have a   
non-zero overlap. The latter property resembles tunnelling due to instantons    
in the usual formulation. Note however that in our picture one did not   
consider minima of the {\it classical} action. The lowest energy states have   
been obtained within the quantum mechanical treatment of the residual  
symmetry consisting of the c-number shifts of an operator.

Implementation of large gauge transformations for the dynamical fermion 
field $\psip(\xmin)$ is based on the commutator  
\beq 
\left[\psip(\xmin),j^+(\ymin)\right] = \psip(\ymin)\gd_a(\xmin - \ymin)
\label{psijcr}
\eeq
which follows from the basic anticommutation relation (\ref{acr}). The unitary 
operator $\hat{F}(\gn)=(\hat{F}_1)^{\gn}$ that implements the phase 
transformation (\ref{zmshift}) is
\beq
\psip(\xmin) \rightarrow \hat{F}(\gn)\psip(\xmin)\hat{F}^{\dagger}(\gn),\;\;\;
\hat{F}(\gn) = \exp{\left[-i{\gp \over L}\gn\intgl\xmin j^+(\xmin)\right]}.  
\eeq
The Hilbert space transforms correspondingly. But since physical states are  
states with zero total charge and the pairs of operators $b_k^{\dagger}d_l^
{\dagger}$, which create these states, are gauge invariant, it is only the 
vacuum state that transforms: 
\beq
\vert 0 \rangle \rightarrow \hat{F}(\gn) \vert 0 \rangle = \exp{\left[
-\gn\sum_{m=1}^M
{(-1)^m \over m}(A^{\dagger}_m - A_m)\right]} \vert 0 \rangle \equiv \vert \gn;
f \rangle .
\label{fermivac}
\eeq
The boson Fock operators $A_m, A^{\dagger}_m$ \cite{EP} 
\bea
A_m & = & \sum_{k=\hlf}^{m-\hlf}d_{m-k}b_k + \sum_{k=\hlf}^N 
\left[b^{\dagger}_k b_{m+k} - d^{\dagger}_k d_{m+k}\right],
\nonumber \\
A^{\dagger}_m & = & \sum_{k=\hlf}^{m-\hlf}b^{\dagger}_{k}d^
{\dagger}_{m-k} + \sum_{k=\hlf}^N
\left[b^{\dagger}_{m+k}b_k - d^{\dagger}_{m+k}d_k\right],
\eea
satisfying $\left[A_m,A_{m^\prime}^{\dagger}\right] = \sqrt{m m^{\prime}}
\gd_{m,m^{\prime}}$ emerge naturally after taking a Fourier transform of  
$j^+(\xmin)$ expressed in terms of fermion modes. This yields  
\beq
j^+(\xmin) = {1 \over{L}}\sum_{m=1}^{M}\left[A_m e^{-{i\over{2}}p^+_m
\xmin} + A^{\dagger}_m e^{{i\over{2}}p^+_m\xmin}\right]
\eeq
as well as the exponential operator in Eq.(\ref{fermivac}). The states   
$\vert \gn;f \rangle$ are not invariant under $\hat{F}_1$: $\vert \gn;f 
\rangle \rightarrow \vert \gn + 1;f \rangle $, in analogy with the 
 Eq.(\ref{psishift}). 
 
To construct the physical vacuum state of the massive Schwinger model, one  
first defines the operator of the full large gauge transformations $\hat{T}_1$ 
as a product of commuting operators $\hat{Z}_1$ and $\hat{F}_1$. The 
requirement of gauge invariance of the physical ground state then leads to the 
$\gt$-vacuum, which is obtained by diagonalization, i.e. by summing the 
degenerate vacuum states $\vert \gn \rangle = \vert \gn;z \rangle \vert \gn;f 
\rangle$ with the appropriate phase factor:
\beq
\vert \theta \rangle = \sum_{\gn=-\infty}^{+\infty}\!\!e^{i\gn\theta}\vert\gn  
\rangle = \sum_{\gn=-\infty}^{+\infty}\!\!e^{i\gn\theta}\left(
\hat{T}_1\right)^\gn \vert 0 
\rangle,\;\;\hat{T}_1 \vert \theta \rangle = e^{-i\theta}\vert \theta \rangle,
\;\;\hat{T}_1 \equiv \hat{Z}_1\hat{F}_1,
\label{theta}
\eeq
$(\vert 0 \rangle$ here denotes both the fermion and gauge boson Fock vacuum). 
Thus we see that the $\theta$-vacuum $\vert \theta \rangle$ is an eigenstate 
of $\hat{T}_1$ with the eigenvalue $\exp(-i\theta)$. In other words, it is 
invariant up to a phase, which is the usual result \cite{LSw,Stroc}.  

The physical meaning of the vacuum angle $\theta$ as the constant background 
electric field \cite{Colm} can be found by a straightforward calculation:   
$\langle \gt \vert \gP_{\A0} \vert \gt \rangle = {e\gt \over{2\gp}}$, where 
the infinite normalization factor $\langle \gt \vert \gt \rangle$ has been 
divided out.

The $\vert \gn \rangle$-vacuum expectation values of $P^-$ are degenerate  
due to gauge invariance of the latter. Subtracting the value (\ref{zmvacen})  
as well as another constant coming from the normal-ordering of the mass term  
\cite{LMlong}, this vacuum expectation value can be set to zero. Then one has  
$\langle \gt \vert P^- \vert \gt \rangle = 0$, while $\langle \gt \vert P^+ 
\vert \gt \rangle = 0$  and $Q\vert \gt \rangle = 0 $ automatically 
\cite{LMlong}.      

Finally, we would like to point out that the fermion component of the 
theta-vacuum (\ref{theta}), described in terms of the exponential of the 
effective boson operators $A_m, A^{\dagger}_m$, introduces a possibility of    
obtaining a non-vanishing fermion condensate in the LF massive Schwinger model.
 
\section{LF Vacuum in Other Gauge Theories}

Let us consider briefly the application of the above ideas to more complicated 
gauge theories. The first example is the two-dimensional $SU(2)$ Yang-Mills 
theory with colour massive fermion field $\Psi_i(x),i=1,2$, in the fundamental 
representation. The gauge field is defined by means of the Pauli matrices  
$\gs^a, a=1,2,3$, as $A^{\gm a}(x) = A^{\gm a}(x) {\gs^a \over 2}$.  

The gauge fixing in the model can be performed analogously to the massive 
Schwinger model by setting  $A^{+a}_n = 0, A^{-a}_0 = 0$. In the finite   
volume, the residual gauge symmetry, represented by constant $SU(2)$ matrices,  
permits to diagonalize $A^+_0$. Consequently, there is only one dynamical 
gauge field ZM for the $SU(2)$ theory, namely $A_0^{+3}={2\gp\over{gL}}\hat{\gz 
}$, where $g$ is the gauge coupling constant. The LF Hamiltonian, which is a 
$SU(2)$ generalization of the expression   
(\ref{lfham}), is invariant under residual large gauge transformations  
\beq
A_0^+ \rightarrow A_0^+ - {2\gp\over{gL}}{\gs^3 \over 2},\;\;\psip^i(\xmin) 
\rightarrow e^{i{\gp\over L}{\gs^3 \over 2}\xmin}\psip^i(\xmin),\;i=1,2.
\eeq
Their implementation in coordinate representation is analogous to the abelian 
case with one important difference \cite{Lenzax,Alex}: in order to correctly 
define the ZM momentum and kinetic energy operators, one has to take into 
account the Jacobian $J(\gz)$, which is induced by the curvature of 
the $SU(2)$ group manifold:
\beq
P^-_{0} = -\hlf{{e^2L}\over{2\gp^2}}{1 \over J}{d\over{d\gz}}J{d \over{d\gz}}, 
\;\hat{\gP}_0 = -i{1 \over{\sqrt{J}}}{d \over{d\gz}}\sqrt{J}= 
-i{d \over {d\gz}}-i\gp\cot\gp\gz,\;J = \sin^2\gp\gz. 
\eeq
The presence of the Jacobian has a profound impact on the structure of the ZM 
vacuum wave functions. Defining again the vacuum state as $(\hat{\gz} + i\hat
{\gP}_0)\Psi_0 = 0$, one finds 
\beq
\Psi_0(\gz) = \gp^{-{1 \over 4}}{e^{-\hlf\gz^2}\over{\vert \sin\gp\gz \vert}}
\rightarrow \Psi_{\gn}(\gz) = e^{-i\gn\hat{\gP}_0}\Psi_0(\gz) = 
\gp^{-{1 \over 4}}{e^{-\hlf(\gz - \gn)^2}\over{\vert \sin\gp\gz \vert}}.
\eeq 
Thus, each wave function is divided into pieces separated by singular points 
at integer multiples of $\gp$ and individual states are just shifted copies   
of $\Psi_0(\gz)$ with no overlap. Consequently, the $\gt$-vacuum cannot be  
constructed \cite{Wit,Engelh}. Further details will be given separately  
\cite{LJSU2}. 

It is rather striking that the generalization of the present approach to the 
vacuum problem for the case of the LF QED(3+1), quantized in the (generalized) 
LC gauge and in a finite volume $-L \le \xmin \le L,\; -L_{\perp} \le x^j\le 
L_{\perp},j=1,2$, appears to be straightforward. The crucial point is that 
in spite of two extra space dimensions, there is still only one dynamical ZM, 
namely $\A0$ (the subscript 0 indicates the  $(\xmin,x^j)$-independent 
component). Indeed, $A_0^-$ can be gauged away (see below) and $A_0^j$ are 
constrained. Proper zero modes, i.e. the gauge field components $a^+, a^-, a^j$
that have $p^+=0,p^j \neq 0$, are not dynamically independent variables either  
\cite{ADQED} in contrast with the situation in the equal-time  
Hamiltonian approach \cite{Lenzqm}. 

Residual gauge transformations, which are the symmetry of the theory even after 
all redundant gauge degrees of freedom have been completely eliminated by the 
gauge-fixing conditions $A_n^+=0, A_0^-=0, \delp a^+ + \partial_j a^j = 0$  
\cite{ADQED}, are characterized by the same gauge function $\gL_\gn$ as in  
the Schwinger model, since constant shifts of constrained $A_0^j$ in  
$j$ directions are not allowed.  
In this way, we are led to consider essentially the same unitary operators 
implementing the residual symmetry as in the Schwinger model. For example, 
defining the dimensionless quantities $\hat{\gz}$ and $\hat{\gp}_0$ by 
\beq
\A0 = {2 \gp \over{eL}}\hat{\gz},\;\;\gP_{\A0} = {1 \over{(2L_{\perp})^2}}
{e \over {2\gp}}\hat{\gp}_0,
\eeq
one again recovers the commutator (\ref{qmcr}), the shift operator $\hat{Z}
(\gn)$, etc.

Before being able to make conclusions about the $\gt$-vacuum of the 
light-front QED(3+1) \cite{Rub}, one needs to better understand the role of 
constrained zero modes. Let us emphasize only one point here: the fermion part  
of the transformed vacuum state acquires again the simple form of 
Eq.(\ref{fermivac}) with 
generalized boson operators $\tilde{A}_m, \tilde{A}_m^\dagger$ ($\gs$ is the 
spin projection and $k_\perp \equiv k^j = \pm 1,\pm 2,\dots$) 
\bea
\tilde{A}_m^\dagger & = & \sum_{k=\hlf}^M \sum_{k_\perp=-M_\perp}^{M_\perp} 
\sum_{\gs=
\pm \hlf}\left[b^{\dagger}_{m+k,k_\perp,\gs}b_{k,k_\perp,\gs} - d^{\dagger}_
{m+k,k_\perp,\gs}d_{k,k_\perp,\gs}\right] \\ 
& + & \sum_{k=\hlf}^{m-\hlf} \sum_{k_\perp = 
-M_\perp}^{M_\perp} \sum_{\gs=\pm \hlf}b^{\dagger}_{k,k_\perp,\gs}d^{\dagger}_
{m-k,-k_\perp,-\gs}.
\eea
The vacua $\vert \gn;f\rangle$ (\ref{fermivac}) with $\tilde{A}_m^\dagger, 
\tilde{A}_m$ as given above satisfy $\langle \gn;f \vert P^+ \vert \gn;f 
\rangle = 0$, $Q\vert \gn;f \rangle = 0$, as should.

\section{Discussion}

The main result of the present work is the demonstration that, despite the 
apparent ``triviality" of the LF vacuum in the sector of normal modes, it is  
possible to recover the necessary vacuum structure of light-front gauge 
theories. The principal elements of the approach were the infrared    
regularization achieved by quantizing in a finite volume and a systematic  
implementation of the residual large gauge symmetry (specific to the 
compactified formulation) in terms of unitary operators. An infinite set 
of non-trivial non-perturbative vacuum states then emerges as the transformed 
``trivial" Fock vacuum. The requirement of gauge invariance 
(as well as of the cluster property \cite{KSus}) of the ground state yields  
the $\gt$-vacuum in the case of the massive Schwinger model.  

Zero-mode aspects of the LF Schwinger model quantized at $x^+ = 0$ have been  
discussed in the literature before \cite{Rgb,Rgf,AD,LM0}. The massive case has 
been studied in \cite{Hara,LJ}. Fermionic aspects of the residual symmetry are  
usually analyzed within the model (rather ad hoc `N-vacua') for the LF 
fermionic vacuum \cite{Rgf,Hara}. Our construction avoids the introduction 
of the Dirac sea in a natural way.  
The new insight is that fermion degrees of freedom are inevitably present in  
the LF ground state -- though outside the usual Fock-state description -- as  
a {\it consequence} of the residual symmetry under large gauge transformations.
It remains to be seen if other non-perturbative features like fermion 
condensate and axial anomaly can be (at least in the continuum limit) 
reproduced correctly in this approach,  
which uses only fields initialized on one characteristic surface. Also, we 
believe that the physics of the massless model will be recovered in the  
$m \rightarrow 0$ limit of the massive theory. 

Furthermore, a possible generalization of the latter to the LF $SU(2)$ gauge 
theory in two dimensions has been suggested. Structure of the vacuum wave 
functions, changed by a presence of the non-trivial Jacobian, indicates that 
no $\gt$-vacuum can be formed in this case, in agreement with previous 
conclusions \cite{Wit,Engelh}. Although the extension of our approach to the 
vacuum problem of a realistic abelian gauge theory, namely QED(3+1),  
appeared to be rather straightforward, difficulties related to the 
renormalization and the presence of non-dynamical zero modes obeying the 
complicated operator constraints \cite{ADQED} are to be expected. On the 
other hand, a more general method  
\cite{Lenzqm,Lenzax,JAR} of elimination of redundant gauge degrees 
of freedom by unitary transformations may become a useful alternative to 
the conventional gauge-fixed formulation of the light-front quantization.   

\section{Acknowledgements} 

I would like to thank A. Harindranath and J. Vary 
for many profitable discussions and the  
International Institute of Theoretical and Applied Physics at the 
Iowa State University for support and hospitality. 
This work has been supported by the Slovak Grant 
Agency and by the NSF grant No. INT-9515511.

\end{document}